\documentclass[10pt]{amsproc}

\usepackage{amsmath}
\usepackage{amssymb}
\usepackage{amsthm}

\theoremstyle{definition}

\newlength{\fuyasu}
\theoremstyle{definition}

\newcommand{\geh}{\mathfrak{g}}
\newcommand{\gehn}{\ensuremath{\mathfrak{g}_n}}

\newcommand{\ot}{\otimes}

\newcommand{\Z}{{\mathbb Z}}

\begin{document}

\title[Bethe ansatz at $q=0$ and periodic box-ball systems]{
Bethe ansatz at $q=0$ and 
periodic box-ball systems
}

\begin{abstract}
A class of periodic soliton cellular automata is introduced 
associated with crystals of non-exceptional quantum affine algebras.
Based on the Bethe ansatz at $q=0$,
we propose explicit formulae for the dynamical period
and the size of certain orbits under the time evolution 
in the $A^{(1)}_n$ case.

\end{abstract}

\author{Atsuo Kuniba}
\address{Institute of Physics, University of Tokyo, Tokyo 153-8902, Japan}
\email{atsuo@gokutan.c.u-tokyo.ac.jp}

\author{Akira Takenouchi}
\address{Institute of Physics, University of Tokyo, Tokyo 153-8902, Japan}
\email{takenouchi@gokutan.c.u-tokyo.ac.jp}
\date{}

\mathversion{bold}
\maketitle
\mathversion{normal}

\section{Introduction}\label{sec:1}

The box-ball system \cite{TS,T} 
is a soliton cellular automaton on a one dimensional lattice.
It is an ultradiscrete integrable system \cite{TTMS} that exhibits 
factorized scattering, 
and has been studied {}from a variety of aspects.
Among them an efficient viewpoint is 
a solvable vertex model in statistical mechanics \cite{B} at $q=0$, 
where the time evolution of the box-ball system is identified with the 
action of a transfer matrix.
It has led to a direct formulation \cite{HHIKTT,FOY}
by the crystal base theory, a theory of quantum group at $q=0$ \cite{K},
and generalizations associated with quantum affine algebras 
\cite{HKT1, HKOTY}.
For some latest developments along this line, 
see \cite{IKO, KOY}.
These studies are based on the idea of 
commuting transfer matrices \cite{B}.
As a method of analyzing solvable lattice models, it is 
complementary to the most efficient technique 
known as the Bethe ansatz \cite{Be}; therefore 
it is natural to seek its application 
to the box-ball system and its generalizations.

The aim of this paper is to extend 
the box-ball system to periodic versions 
and launch a Bethe ansatz approach to them.
For non-exceptional affine Lie algebra $\gehn$,  
we construct a periodic ultradiscrete dynamical system that tends to 
the $\gehn$ automaton \cite{HKT1} in an infinite lattice limit.
Here is an example of the time evolution pattern  
for $\geh_n = A^{(1)}_2$:
\noindent
\begin{center}
$t=0:\quad 1\;\;1\;\;2\;\;1\;\;3\;\;2$ \\ \vspace{-0.07cm}

$t=1:\quad 3\;\;2\;\;1\;\;2\;\;1\;\;1$ \\ \vspace{-0.07cm}

$t=2:\quad 1\;\;1\;\;3\;\;1\;\;2\;\;2$ \\ \vspace{-0.07cm}

$t=3:\quad 2\;\;2\;\;1\;\;3\;\;1\;\;1$ \\ \vspace{-0.07cm}

$t=4:\quad 1\;\;1\;\;2\;\;2\;\;3\;\;1$ \\ \vspace{-0.07cm}

$t=5:\quad 2\;\;1\;\;1\;\;1\;\;2\;\;3$ \\ \vspace{-0.07cm}

$t=6:\quad 1\;\;3\;\;2\;\;1\;\;1\;\;2$ \\ \vspace{-0.07cm}

$t=7:\quad 2\;\;1\;\;1\;\;3\;\;2\;\;1$ \\ \vspace{-0.07cm}

$t=8:\quad 1\;\;2\;\;2\;\;1\;\;1\;\;3$ \\ \vspace{-0.07cm}

$t=9:\quad 3\;\;1\;\;1\;\;2\;\;2\;\;1$ \\ \vspace{-0.07cm}

$\!\!t=\!10:\quad 2\;\;3\;\;1\;\;1\;\;1\;\;2$ \\ \vspace{-0.07cm}

$\!\!t=\!11:\quad 1\;\;2\;\;3\;\;2\;\;1\;\;1$ \\ \vspace{-0.07cm}

$\!\!t=\!12:\quad 1\;\;1\;\;2\;\;1\;\;3\;\;2$
\end{center}

\noindent
Regarding the letter $1$ as background, 
one observes two solitons 
proceeding cyclically to the right
with velocity $=$ amplitude equal to 2 and 1.
They repeat collisions (or overtaking) under which the reactions 
$32 \times 2 \rightarrow 3 \times 22$ and 
$22 \times 3 \rightarrow 2 \times 32$ take place.
Behind such dynamics there underlines a solvable vertex model 
at $q=0$, where only some selected configurations
have non-zero Boltzmann weights and the transfer matrix
yields a deterministic evolution of the spins on one row to another.
For instance, the transition from $t=0$ to $t=2$ states 
has been determined from the configuration in figure 1 on 
a two-dimensional square lattice:

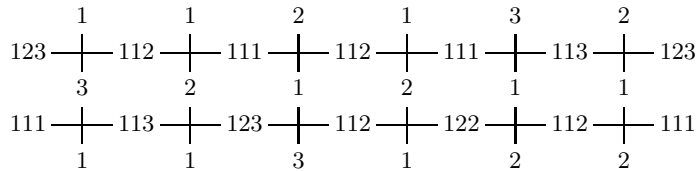
\begin{figure}[h]
\unitlength 0.8mm
{\small
\begin{picture}(110,30)(-6,-19)
\multiput(0,-12)(0,12){2}{
\multiput(0,0)(18,0){6}{\line(1,0){10}}
\multiput(5,-3)(18,0){6}{\line(0,1){6}}
}

\put(4,5){1}\put(22,5){1}\put(40,5){2}
\put(58,5){1}\put(76,5){3}\put(94,5){2}

\put(-7,-1){123}
\put(11,-1){112}
\put(29,-1){111}
\put(47,-1){112}
\put(65,-1){111}
\put(83,-1){113}
\put(101,-1){123}

\put(4,-7){3}\put(22,-7){2}\put(40,-7){1}
\put(58,-7){2}\put(76,-7){1}\put(94,-7){1}

\put(-7,-13){111}
\put(11,-13){113}
\put(29,-13){123}
\put(47,-13){112}
\put(65,-13){122}
\put(83,-13){112}
\put(101,-13){111}

\put(4,-19){1}\put(22,-19){1}\put(40,-19){3}
\put(58,-19){1}\put(76,-19){2}\put(94,-19){2}

\end{picture}
}
\caption{A vertex configuration}\label{fig}
\end{figure}
This is a configuration of the fusion 
$U_q(A^{(1)}_2)$ vertex model 
that survives at $q=0$.
In the terminology of quantum inverse scattering method \cite{STF}, 
the quantum space on vertical lines carries the 
fundamental representation ($1, 2$ or $3$) of 
$A_2 = sl_3$  and 
the auxiliary space on horizontal lines does the 
three-fold symmetric tensor representation
($111, 112, \ldots, 333$).
The automaton states live on the vertical lines.
The dynamics is governed by {\em combinatorial $R$}, which
is the quantum $R$ matrix at $q=0$ 
specified by local configurations round a vertex.
The states on horizontal lines are so chosen that 
they become equal at the both ends 
reflecting the periodic boundary condition.

In this paper we introduce analogous 
periodic automata for any non-exceptional 
affine Lie algebra $\geh_n$ 
based on the factorization of the 
combinatorial $R$ \cite{HKT2}.
They may be viewed as the system of particles 
that undergo pair creation and 
annihilation though the collisions. 
Moreover we exploit how 
the Bethe ansatz at $q=0$ \cite{KN} yields 
the dynamical period and size of certain orbits.
For instance in the above time evolution pattern, 
$t=0$ and $t=12$ states are identical, hence 
the dynamical period is $12$.
We propose the general formula 
(\ref{lcm}) for the dynamical period in $A^{(1)}_n$ case,
which indeed predicts $12$ in the above example.
It is expressed as a least common multiple 
of the rational numbers 
arising from Bethe eigenvalues at $q=0$.

In \cite{KN}, the Bethe equation is linearized into 
the string centre equation and an explicit character formula 
(\ref{eq:comp}) 
has been established by counting off-diagonal solutions to the 
string centre equation. 
It is a version of fermionic formulae
and is called the combinatorial completeness 
of the Bethe ansatz at $q=0$. 
In (\ref{eq:omega}) we 
relate each summand (\ref{eq:R}) 
in the character formula 
to the size of a certain orbit 
under the time evolution.
Such a result will be useful to study the entropy of the automata.

The formulae for the dynamical period (\ref{lcm}) and 
the orbit size (\ref{eq:omega}) 
are novel applications of the 
Bethe ansatz to ultradiscrete integrable systems.
Upon identification of strings in the Bethe ansatz 
with solitons in the automata, they 
reproduce the expressions in \cite{YYT} 
for $A^{(1)}_1$ with $l= \infty$.
In our approach, we also use 
the combinatorial Bethe ansatz at $q=1$ 
\cite{KKR,KR}, namely, rigged configurations and 
their bijective correspondence with automaton highest states.
In this terminology, it is the configuration that plays the role of 
the conserved quantity, 
which is an analogous feature to the infinite system \cite{KOTY}.
It is an interesting problem to synthesize the 
combinatorial Bethe ans{\" a}tze at $q=1$ and $q=0$, 
which will provide a unified perspective on the  
automata on the infinite and the periodic lattices.

The paper is arranged as follows. 
In section \ref{sec:2}, a periodic $\gehn$ automaton is introduced.
It tends to that in \cite{HKT1} in an infinite lattice limit and 
includes the one in \cite{YT,MIT} as a $\gehn = A^{(1)}_n$ case.
In section \ref{sec:3}, the Bethe eigenvalues are investigated at $q=0$.
In section \ref{sec:4}, the dynamical period of the periodic automata
is related to the Bethe eigenvalue studied in section \ref{sec:3}.
In section \ref{sec:5},  sizes of certain orbits are
related to the character formula in \cite{KN}.
In the last two sections conjectures are presented 
with compelling experimental data.
The states treated there are time evolutions of the 
highest ones.
The classes of time evolutions being considered in sections 
\ref{sec:4} and \ref{sec:5} are different.
In fact, the latter is wider containing the former; 
therefore, the `period' in section \ref{sec:4} is a notion 
different {}from the `size of orbit' in section \ref{sec:5}.
A more unified framework 
including a treatment of non-highest states
will be presented elsewhere.
The last table in section \ref{sec:4} is a preliminary 
report on $D^{(1)}_4$.
For standard notations and facts in the crystal base theory, 
we refer to \cite{K,KKM,KMN}.

\section{Periodic $\gehn$ automaton}\label{sec:2}
Let $U_q(\gehn)$ be the quantum affine algebra 
associated with non-exceptional 
$\gehn = A^{(1)}_n$, $A^{(2)}_{2n-1}$, $A^{(2)}_{2n}$,
$B^{(1)}_n$, $C^{(1)}_n$, $D^{(1)}_n$ and $D^{(2)}_{n+1}$.
Denote by $B_l$ the crystal of the $l$-fold symmetric fusion 
of the vector representation of $U_q(\gehn)$ \cite{KKM}.
We are going to introduce a dynamical system on the 
finite tensor product $B:= 
B_{l_1} \otimes B_{l_2} \otimes \cdots \otimes B_{l_L}$.
An element of $B$ will be called a {\em path}.
The representative time evolution is given by
\begin{equation}\label{eq:tinf}
T_\infty = \sigma_B
S_{i_{d}}\cdots S_{i_{2}}S_{i_{1}}.
\end{equation}
Here $S_i$ is the Weyl group operator \cite{K} and 
$\sigma_B = \overbrace{\sigma \otimes \cdots \otimes \sigma}^L$
with each $\sigma$ acting on the components $B_{l_i}$ individually 
according to Table \ref{tab:1}.
For example for the element $11245 \in B_5$ of $A^{(1)}_4$ 
represented by the semistandard tableau, 
one has $\sigma(11245) = 13455$.
See \cite{HKT2} section 2 for the notation in the other algebras.
We call the dynamical system on $B$ with the time evolution 
(\ref{eq:tinf}) the periodic $\gehn$ automaton.
In case $B$ is of the form $B = B_1^{\otimes L}$, it will be 
called the basic periodic $\gehn$ automaton.

\begin{table}[h]
\caption{The data $\sigma, d$ and $i_k$}\label{tab:1}
\begin{center}%
\small{%
\begin{tabular}{l|c|c|c|c}
$\gehn$ & $\sigma$ & $d$ & $i_d, \ldots, i_1$ & $W$  \\ \hline
$A^{(1)}_n$ & $a \mapsto a-1$ & $n$ & $2,3,\ldots, n\!-\!1,n,0$  
& $W(A_{n\!-\!1})$ \\
$A^{(2)}_{2n\!-\!1}$ & $1 \leftrightarrow \overline{1}$ & $2n\!-\!1$
& $0,2,\ldots, n\!-\!1,n,n\!-\!1,\ldots,2, 0$ 
& $W(BC_{n\!-\!1})$ \\
$A^{(2)}_{2n}$ & $ id$ & $2n$
& $1, 2, \ldots, n\!-\!1,n,n\!-\!1,\ldots,2,1,0$ 
&$W(BC_{n\!-\!1})$ \\
$B^{(1)}_{n}$ & $1 \leftrightarrow \overline{1}$ & $2n\!-\!1$
&$0,2,\ldots, n\!-\!1,n,n\!-\!1,\ldots,2, 0$ 
& $W(BC_{n\!-\!1})$ \\
$C^{(1)}_{n}$ & $id$ & $2n$
& $1, 2, \ldots, n\!-\!1,n,n\!-\!1,\ldots,2,1,0$ 
& $W(BC_{n\!-\!1})$ \\
$D^{(1)}_{n}$ & $1 \leftrightarrow \overline{1},
n \leftrightarrow \overline{n}$ & $2n\!-\!2$
& $0,2, \ldots, n\!-\!2,{n\!-\!1, n \brace n, n\!-\!1}, n\!-\!2,\ldots, 2,0$ 
& $W(D_{n\!-\!1})$ \\
$D^{(2)}_{n\!+\!1}$ & $id$ & $2n$
& $1, 2, \ldots, n\!-\!1,n,n\!-\!1,\ldots,2,1,0$ 
&$W(BC_{n\!-\!1})$ \\
\end{tabular}
}
\end{center}
\end{table}

Let us illustrate (\ref{eq:tinf}) along a 
$\geh_n = A^{(1)}_2$ example.
The time evolution 
$T_\infty = \sigma_BS_2S_0$ of the $t=0$ path 
$1 1 2 1 3 2 \in B_1^{\otimes 6}$ into 
$3 2 1 2 1 1$ at $t=1$ in section \ref{sec:1} is computed as
\begin{equation}\label{eq:texample}
\underset{2-{\rm signaure}}{{\underset{0-{\rm signature}}
{\phantom{1-signature}}}}
\underset{ }{\underset{-}{1}} \; 
\underset{ }{\underset{-}{1}} \; 
\underset{(+}{\underset{ }{2}} \,
\underset{ }{\underset{-}{1}} \,
\underset{-)}{\underset{+}{3}} \; 
\underset{+}{\underset{ }{2}} \; 
\overset{S_0}{\mapsto} 
\underset{ }{\underset{-}{1}} \; 
\underset{-}{\underset{+}{3}} \; 
\underset{(+ }{\underset{ }{2}}
\underset{\,-)}{\underset{+}{3}} \; 
\underset{-}{\underset{+}{3}} \; 
\underset{+}{\underset{ }{2}} \; 
\overset{S_2}{\mapsto} 
\underset{ }{\underset{-}{1}} \; 
\underset{-}{\underset{+}{3}} \; 
\underset{(+}{\underset{ }{2}} 
\underset{\,-)}{\underset{+}{3}} \, 
\underset{+}{\underset{ }{2}} \; 
\underset{+}{\underset{ }{2}} \; 
\overset{\sigma_B}{\mapsto} 
\underset{ }{\underset{ }{3}} \;
\underset{ }{\underset{ }{2}} \;
\underset{ }{\underset{ }{1}} \; 
\underset{ }{\underset{ }{2}} \; 
\underset{ }{\underset{ }{1}} \; 
\underset{ }{\underset{ }{1}}.
\end{equation}
For the first three paths, we have exhibited the 
$0$-signature and $2$-signature. In general, 
the $i$-signature of an element $a$ in the 
$A^{(1)}_2$ crystal $B_1 = \{1, 2, 3\}$
is the symbol $+$ if $a=i$, $-$ if $a=i+1$ mod 3 and none otherwise.
{}From the array of $i$-signatures, one eliminates 
the pair $+-$ (not $-+$) successively to 
finally reach the pattern 
$\overbrace{-\cdots -}^\alpha\overbrace{+ \cdots +}^\beta$
called the reduced $i$-signature.
Then the action of $S_i$ is unambiguously defined as 
the interchange 
$\overbrace{-\cdots -}^\alpha\overbrace{+ \cdots +}^\beta
\mapsto 
\overbrace{-\cdots -}^\beta\overbrace{+ \cdots +}^\alpha$
on the reduced $i$-signature.
In (\ref{eq:texample}), we have shown the elimination of 
the $+-$ pairs by parentheses.
By the very same rule, the 
Weyl group operators in general $\geh_n$ and 
$B = B_{l_1} \otimes \cdots \otimes B_{l_L}$ can be 
computed using the necessary data on $B_l$ in \cite{KKM}. 

One may wonder the relation between the two derivations 
of the time evolution 
$1 1 2 1 3 2 (t\!=\!0) \mapsto 
3 2 1 2 1 1 (t\!=\!1)$,  one as (\ref{eq:texample}) and 
the other as in Figure \ref{fig} in section \ref{sec:1}.
Let us clarify it by explaining the origin of (\ref{eq:tinf}).
Recall that the automata in the infinite system 
\cite{HKT1, HHIKTT, FOY, HKOTY} have the set of states 
$\cdots \otimes B_{l_i} \otimes B_{l_{i+1}} \otimes \cdots$ with 
the boundary condition that the sufficiently distant local states 
are the highest element $u_{l_i} = (1^{l_i}) \in B_{l_i}$.
The commuting family of time evolutions 
$T_l\; (l \in \Z_{\ge 1})$ is induced by the relation 
\begin{equation}\label{eq:tl}
\begin{split}
&B_l \otimes 
(\cdots \otimes B_{l_i} \otimes B_{l_{i+1}} \otimes \cdots) 
\simeq 
(\cdots \otimes B_{l_i} \otimes B_{l_{i+1}} \otimes \cdots)\otimes B_l\\
&\;\qquad \qquad u_l \otimes p \qquad \qquad \qquad \simeq 
\qquad \qquad \quad T_l(p) \otimes u_l
\end{split}
\end{equation}
under the isomorphism of crystals.
It was proved in \cite{HKT2} that $T_l$ with sufficiently large $l$ 
is factorized as (\ref{eq:tinf}), where 
all the $S_i$ actually act as ${\tilde e}^{\infty}_i$.
In this sense (\ref{eq:tinf}) is a natural analogue of the $T_\infty$ 
in the infinite system, which corresponds to the 
limit of the periodic $\gehn$ automaton 
when the system size $L$ grows to infinity 
under the above-mentioned boundary condition.
The product (\ref{eq:tinf}) 
is a translation in the extended affine Weyl group.
The indices $i_k$ in Table \ref{tab:1} 
are equal to $i_{k+j}$ in \cite{HKT2} for some $j$.
They have been chosen so that the tableau letter 
representing the background or `empty box' to becomes 1.

Let us comment on 
the analogue of the time evolution $T_l$ with finite $l$ 
on our periodic $\geh_n$ automaton.
A natural idea is to define it by 
an analogue of the relation (\ref{eq:tl}) as
$v_l  \otimes p \simeq p' \otimes v_l$, where 
$v_l \in B_l$ is not necessarily the highest element $u_l$ 
in general.
If such a $v_l$ exists and $p'$ is unique even when 
$v_l$ is not unique, we set $T_l(p)=p'$ and say that 
$T_l(p)$ exists.
$T_l(p)$ does not always exist.
For instance in $A^{(1)}_n$ case, 
$v_1$ does not exist for $L=l=1, p = 12 \in B=B_2$, and 
$p'$ is not unique for 
$L=2, l=1, p = 12 \otimes 12 \in B_2\otimes B_2$.
See section \ref{sec:5} for more arguments.
On the other hand 
for $l$ sufficiently large, we expect that 
$T_l(p)$ exists. In fact the following assertion is valid.

\vspace{0.2cm}\noindent
{\bf Theorem}. 
Let $\gehn=A^{(1)}_n$ (hence $d=n$).  Pick any element $p \in B$ such that 
\begin{equation}\label{eq:cond}
\varphi_{i_k}(S_{i_{k-1}}\cdots S_{i_1}(p)) \le 
\varepsilon_{i_k}(S_{i_{k-1}}\cdots S_{i_1}(p)) 
\quad \hbox{for } 1 \le k \le d.
\end{equation}
Set $v_l=(x_1,\ldots, x_{n+1})$. Here the number 
$x_i\in \Z_{\ge 0}\, (i \in \Z_{n+1})$ of the letter $i$ 
in the semistandard tableau on length $l$ row is determined by
$x_{i_k} = \varphi_{i_k}(S_{i_{k-1}}\cdots S_{i_1}(p))$ 
for $1 \le k \le n$ and
$x_1 + \cdots + x_{n+1} = l$, which is possible for $l$ large.
Then for sufficiently large $l$, the relation 
\begin{equation}\label{eq:v}
v_l \otimes p \simeq T_{\infty}(p) \otimes v_l
\end{equation}
holds under the isomorphism of crystals 
$B_l \ot B \simeq B \otimes B_l$ with $T_\infty$ given by (\ref{eq:tinf}).

\vspace{0.2cm}
The condition (\ref{eq:cond}) stated in an intrinsic manner, 
is actually a simple 
postulate that among $\{1, \ldots, n\!+\!1\}$,  
the letter 1 should be no less than any other ones in the 
semistandard tableaux consisting of $p$.
On the paths in Figure \ref{fig} 
($112132, 321211, 113122 \in B^{\otimes 6}_1$), 
one has $T_3 = T_\infty$. 
A similar theorem is valid also for $D^{(1)}_n$.

The time evolution $T_\infty$ (\ref{eq:tinf}) 
commutes with several operators acting on $B$, which form the 
symmetry of ($T_\infty$ flow of) our periodic $\gehn$ automaton.
By using (\ref{eq:tinf}) and 
$S_i\sigma_B = \sigma_BS_{\sigma^{-1}(i)}$ (see \cite{HKT2}), 
it is easy to check that $T_\infty S_i = S_iT_\infty$ 
for $i \neq 0, 1$.
The symmetry operators or 
`B\"acklund transformations' $\{S_i \mid 2 \le i \le n\}$ form
a classical Weyl group listed in the rightmost column of 
Table \ref{tab:1}.
This is a smaller symmetry compared with the 
$U_q(\overline{\mathfrak g}_{n-1})$-invariance 
in the case of the infinite system \cite{HKOTY}. 

Here is an example of the time evolution (downward) 
in the periodic $A^{(1)}_3$ automaton 
with $B = B_3\otimes B_1 \otimes B_1 \otimes B_1
\otimes B_1 \otimes B_2 \otimes B_1$.
At each time step, the paths connected by 
the Weyl group actions $S_2$ and $S_3$ are shown, 
forming commutative diagrams.

\begin{center}
$133 \cdot 4 \cdot 1 \cdot 3 \cdot 4 \cdot 12 \cdot 4 
\;\stackrel{S_2}{\mapsto}\;
 123 \cdot 4 \cdot 1 \cdot 2 \cdot 4 \cdot 12 \cdot 4 
\;\stackrel{S_3}{\mapsto}\;
 123 \cdot 4 \cdot 1 \cdot 2 \cdot 3 \cdot 12 \cdot 3 $\\

$124 \cdot 3 \cdot 4 \cdot 1 \cdot 3 \cdot 14 \cdot 3
\;\;\phantom{\stackrel{S_2}{\mapsto}}\;\;
 124 \cdot 3 \cdot 4 \cdot 1 \cdot 2 \cdot 14 \cdot 2 
\;\;\phantom{\stackrel{S_2}{\mapsto}}\;\;
 123 \cdot 3 \cdot 4 \cdot 1 \cdot 2 \cdot 13 \cdot 2 $\\

$134 \cdot 2 \cdot 3 \cdot 4 \cdot 1 \cdot 34 \cdot 1
\;\;\phantom{\stackrel{S_2}{\mapsto}}\;\;
124 \cdot 2 \cdot 3 \cdot 4 \cdot 1 \cdot 24 \cdot 1
\;\;\phantom{\stackrel{S_2}{\mapsto}}\;\;
123 \cdot 2 \cdot 3 \cdot 4 \cdot 1 \cdot 23 \cdot 1$\\

$134 \cdot 1 \cdot 2 \cdot 2 \cdot 4 \cdot 13 \cdot 4
\;\;\phantom{\stackrel{S_2}{\mapsto}}\;\;
124 \cdot 1 \cdot 2 \cdot 3 \cdot 4 \cdot 12 \cdot 4
\;\;\phantom{\stackrel{S_2}{\mapsto}}\;\;
123 \cdot 1 \cdot 2 \cdot 3 \cdot 4 \cdot 12 \cdot 3$

\end{center}
A similar example {}from the basic periodic $D^{(1)}_4$ automaton 
with $B = B^{\otimes 12}_1$.
\begin{center}

$\bar{2}\; 2\; 2\; 2\;1\;1\;1\;1\;4\;\bar{2}\;1\;1 
\;\stackrel{S_2}{\mapsto}\; 
\bar{2}\;3\;2\;2\;1\;1\;1\;1\;4\;\bar{2}\;1\;1
\;\stackrel{S_4}{\mapsto}\; 
\bar{2}\;\bar{4}\;2\;2\;1\;1\;1\;1\;\bar{3}\;\bar{2}\;1\;1$\\

$1\;1\;1\;1\;\bar{2}\;2\;2\;2\;1\;4\;\bar{2}\;1
\;\;\phantom{\stackrel{S_2}{\mapsto}}\;\;
1\;1\;1\;1\;\bar{2}\;3\;2\;2\;1\;4\;\bar{2}\;1
\;\;\phantom{\stackrel{S_4}{\mapsto}}\;\;
1\;1\;1\;1\;\bar{2}\;\bar{4}\;2\;2\;1\;\bar{3}\;\bar{2}\;1$\\

$2\;1\;1\;1\;1\;1\;1\;1\;\bar{2}\;2\;4\;\bar{1}
\;\;\phantom{\stackrel{S_2}{\mapsto}}\;\; 
2\;1\;1\;1\;1\;1\;1\;1\;\bar{2}\;3\;4\;\bar{1}
\;\;\phantom{\stackrel{S_4}{\mapsto}}\;\;
2\;1\;1\;1\;1\;1\;1\;1\;\bar{2}\;\bar{4}\;\bar{3}\;\bar{1}$\\

$2\;\bar{2}\;\bar{3}\;4\;2\;1\;1\;1\;1\;1\;3\;1
\;\;\phantom{\stackrel{S_2}{\mapsto}}\;\;
2\;\bar{2}\;\bar{2}\;4\;2\;1\;1\;1\;1\;1\;3\;1
\;\;\phantom{\stackrel{S_4}{\mapsto}}\;\;
2\;\bar{2}\;\bar{2}\;\bar{3}\;2\;1\;1\;1\;1\;1\;\bar{4}\;1$\\

$1\;2\;1\;1\;1\;\bar{2}\;\bar{3}\;4\;2\;1\;1\;3
\;\;\phantom{\stackrel{S_2}{\mapsto}}\;\;
1\;2\;1\;1\;1\;\bar{2}\;\bar{2}\;4\;2\;1\;1\;3
\;\;\phantom{\stackrel{S_4}{\mapsto}}\;\;
1\;2\;1\;1\;1\;\bar{2}\;\bar{2}\;\bar{3}\;2\;1\;1\;\bar{4}$\\

$4\;1\;3\;2\;1\;1\;1\;1\;1\;\bar{2}\;\bar{3}\;2
\;\;\phantom{\stackrel{S_2}{\mapsto}}\;\;
4\;1\;3\;2\;1\;1\;1\;1\;1\;\bar{2}\;\bar{2}\;2
\;\;\phantom{\stackrel{S_4}{\mapsto}}\;\;
\bar{3}\;1\;\bar{4}\;2\;1\;1\;1\;1\;1\;\bar{2}\;\bar{2}\;2$\\

$3\;\bar{3}\;1\;1\;\bar{3}\;4\;3\;2\;1\;1\;1\;1
\;\;\phantom{\stackrel{S_2}{\mapsto}}\;\; 
3\;\bar{2}\;1\;1\;\bar{3}\;4\;3\;2\;1\;1\;1\;1
\;\;\phantom{\stackrel{S_4}{\mapsto}}\;\;
3\;\bar{2}\;1\;1\;\bar{3}\;\bar{3}\;\bar{4}\;2\;1\;1\;1\;1$\\

$1\;3\;\bar{3}\;1\;1\;1\;1\;1\;\bar{3}\;4\;3\;2
\;\;\phantom{\stackrel{S_2}{\mapsto}}\;\;
1\;3\;\bar{2}\;1\;1\;1\;1\;1\;\bar{3}\;4\;3\;2
\;\;\phantom{\stackrel{S_4}{\mapsto}}\;\;
1\;3\;\bar{2}\;1\;1\;1\;1\;1\;\bar{3}\;\bar{3}\;\bar{4}\;2$

\end{center}

Let us remark on another family of maps on $B$, 
which may also be regarded as time evolutions.
For $A^{(1)}_n$ it is a dual of (\ref{eq:tinf})  (cf. \cite{KNY}).
Consider 
the maps ${\mathcal T}_1, \ldots, {\mathcal T}_L:  B \rightarrow B$ defined by
\begin{equation}\label{eq:yang}
\begin{split}
{\mathcal T}_i &= R_{i\!-\!1 \,i} \cdots R_{23}R_{12}P_i 
R_{L\!-\!1\, L}\cdots R_{i\!+\!1\, i\!+\!2} R_{i\, i\!+\!1},\\
P_i &: B^{\vee i} \otimes B_{l_i} \rightarrow 
B_{l_i} \otimes B^{\vee i}\\
& \qquad p \otimes b \quad\mapsto \;\;b \otimes p.
\end{split}
\end{equation}
Here $R_{k\, k\!+\!1}$ is the combinatorial $R$ that exchanges the 
$k$-th and $(k\!+\!1)$-th component{} from the left, 
and $B^{\vee i} = 
B_{l_1} \otimes \cdot\cdot \overset{\stackrel{i}{\vee}}{} \cdot \cdot \otimes B_{l_L}$ 
is the $B$ without the component $B_{l_i}$.
It is an observation going back to \cite{Y} that 
the Yang-Baxter equation and the inversion relation of $R$ lead to
the commuting family 
${\mathcal T}_i{\mathcal T}_j = {\mathcal T}_j{\mathcal T}_i$.
Note that 
${\mathcal T}_i = {\mathcal T}_{i+1}$ when $l_i = l_{i+1}$.

\section{Bethe eigenvalues at $q=0$}\label{sec:3}

In this section we exclusively consider the simply laced $\gehn$.
Eigenvalues of row transfer matrices in 
trigonometric vertex models are 
given by the analytic Bethe ansatz \cite{R,KS}.
In the present case, the relevant quantity is the top term
of $\Lambda^{(1)}_l(u)$ ((2.12) in \cite{KS} modified with 
a parameter $\hbar$ to fit the notation here):
\begin{equation}\label{aba}
\frac{Q_1(u-l\hbar)}{Q_1(u+l\hbar)}
\end{equation}
at the shift (or Hamiltonian) point $u=0$.
Here $Q_1(u)= \prod_k\sinh\pi(u-\sqrt{-1}u^{(1)}_k)$, where 
$\{u^{(a)}_j\}$ are to satisfy 
the Bethe equation eq.(2.1) in \cite{KN}.
For the string solution 
(\cite{KN} Definition 2.3), (\ref{aba}) with $u=0$ tends to  
\begin{equation}\label{la0}
\Lambda_l := \prod_{j\alpha}(-z^{(1)}_{j\alpha})^{\min(j,l)}
\end{equation}
in the limit $q = \exp(-2\pi\hbar) \rightarrow 0$.
Here $z^{(a)}_{j\alpha}$ 
is the center of the $\alpha$-th string 
having color $a$ and length $j$.
Denote by $m^{(a)}_j$ the number of such strings.
The product in (\ref{la0}) is taken over 
$j\in \Z_{\ge 1}$ and $1 \le \alpha \le m^{(1)}_j$.
At $q=0$ the Bethe equation becomes 
the string centre equation (\cite{KN} (2.36)):
\begin{equation}
\prod_{(b,k)\in H} \prod_{\beta = 1}^{m^{(b)}_k}
(z^{(b)}_{k\beta})^{A_{aj\alpha,bk\beta}} =
 (-1)^{p^{(a)}_j+m^{(a)}_j+1},\label{sce0}
\end{equation}
where 
$H := \{(a,j)\mid 1 \le a \le n, j\in \Z_{\ge 1}, 
m^{(a)}_j>0\}$ (denoted by $H'$ in \cite{KN}).
$A_{aj\alpha,bk\beta}$ and $p^{(a)}_j$ are defined by
\begin{align}
A_{aj\alpha,bk\beta} &= 
\delta_{ab}\delta_{j k}\delta_{\alpha \beta}
(p^{(a)}_j+m^{(a)}_j) +
C_{ab}\min(j,k) - \delta_{ab}\delta_{j k},\\
p^{(a)}_j &=\sum_{k\ge 1}\min(j,k)\nu^{(a)}_k
- \sum_{(b,k) \in H}C_{ab}\min(j,k)m^{(b)}_k,\label{eq:vacancy}
\end{align}
where $(C_{ab})_{1 \le a,b \le n}$ 
is the Cartan matrix of the classical part of $\gehn$.
The integer $\nu^{(a)}_k$ is the number of 
the Kirillov-Reshetikhin modules $W^{(a)}_k$ contained in the 
quantum space on which the transfer matrices act.
In our case, the crystal of the quantum space is taken as
$B= B_{l_1} \otimes \cdots \otimes B_{l_L}$ in section \ref{sec:2}, 
hence 
$\nu^{(a)}_k = \delta_{a 1}(\delta_{k l_1} + \cdots + \delta_{k l_L})$.
To avoid a notational complexity 
we temporally abbreviate the triple indices $a j \alpha$ 
to $j$, $b k \beta$ to $k$ and accordingly $z^{(b)}_{k\beta}$ 
to $z_k$ etc.
Then (\ref{la0}) reads
\begin{equation}
\Lambda_l = \prod_{k}(-z_k)^{\rho_{k}},\label{la}
\end{equation}
where $\rho_k$ is actually dependent on $l$, and 
given by $\rho_{k} = \delta_{b 1}\min(k,l)$
for $k$ corresponding to $bk\beta$.
The string centre equation (\ref{sce0}) is written as
\begin{equation}
\prod_k (-z_k)^{A_{j,k}} = (-1)^{s_j}\label{sce}
\end{equation}
for some integer $s_j$.
Note that $A_{j,k}=A_{k,j}$.
Suppose that the $q=0$ eigenvalue (\ref{la}) satisfies 
$\Lambda_l^{{\mathcal P}_l} = \pm 1$ for generic solutions to 
the string centre equation (\ref{sce}).
It means that there exist integers $r_j$ such that 
$\sum_jr_jA_{j,k} = {\mathcal P}_l\rho_{k}$, or equivalently
\begin{equation}\label{linear}
r_j = {\mathcal P_l}\frac{\det A[j]}{\det A},
\end{equation}
where $A[j]$ denotes the matrix $A=(A_{j,k})$ 
with its $j$-th column replaced by 
$^t(\rho_{1},\rho_{2},\ldots)$.
In view of the condition $\forall r_j \in \Z$, 
the minimum integer value allowed for ${\mathcal P}_l$ is
\begin{equation}\label{lcm0}
{\mathcal P}_l = {\rm LCM}\Bigl(1, \, 
\bigcup_k{}^\prime{}\frac{\det A}{\det A[k]} \Bigr),
\end{equation}
where LCM stands for the least common multiple 
and $\cup_k^\prime$ means the union over those $k$ such that 
$A[k] \neq 0$.
The determinants here can be 
simplified by elementary transformations (cf. \cite{KN} (3.9)).
The result is expressed in terms of determinants of 
matrices with indices in $H$:
\begin{equation}\label{lcm}
{\mathcal P}_l = {\rm LCM}\Bigl(1, 
\bigcup_{(b,k) \in H}\!\!\!\!\!{}^\prime\;\;\frac{\det F}{\det F[b,k]} \Bigr),
\end{equation}
where the matrix $F=(F_{aj, bk})_{(a,j), (b,k) \in H}$ is defined by
\begin{equation}\label{F}
F_{aj, bk} = \delta_{ab}\delta_{jk}p^{(a)}_j + C_{a b}\min(j,k)m^{(b)}_k.
\end{equation}
The matrix $F[b,k]$ is obtained {}from $F$ by replacing its 
$(b,k)$-th column as
\begin{equation}
F[b,k]_{aj,cm} = \begin{cases}
F_{aj,cm} & (c,m) \neq (b,k),\\
\delta_{a 1}\min(j,l) & (c,m)=(b,k).
\end{cases}
\end{equation}
The union in (\ref{lcm}) is taken over those $(b,k)$ such that 
$\det F[b,k] \neq 0$.
See also the remark before Conjecture 1 in section \ref{sec:4}.

The LCM in (\ref{lcm}) can further be simplified 
when $\gehn = A^{(1)}_1$ and $\nu^{(1)}_j = L\delta_{j1}$.
We write $m^{(1)}_j, p^{(1)}_j, F[1,k]$ just as 
$m_j, p_j, F[k]$ and parameterize the set
$H = \{ j \in \Z_{\ge 1} \mid m_j > 0\}$ as  
$H = \{(0 <) J_1 < \cdots < J_s \}$. 
The matrix $F[k]$ is obtained by replacing the 
$k$-th column of $F$ by $^t(\min(J_1,l), \min(J_2,l),\ldots)$. 
A direct calculation leads to 
\begin{align}
\det F &= p_{J_0}p_{J_1}\cdots p_{J_{s-1}},\label{Fsl2}\\
\det F[k+1]-\det F[k] &= \frac{p_{J_0}p_{J_1}\cdots p_{J_{s-1}}
p_{i_s}(i_{k+1}-i_k)}
{p_{i_{k+1}}p_{i_k}},\quad 0 \le k \le s-1,\label{FFsl2}
\end{align}
where we have set $i_k = \min(J_k,l)$, $J_0=0$, 
$i_0 = 0$, $p_0 = L$ 
and $F[0]=0$.
($i_k$ here is not related to those in Table \ref{tab:1}.)
Substituting (\ref{Fsl2}) into (\ref{lcm}) and using the 
elementary property of LCM, we find
\begin{equation}\label{eq:lcmsl2}
\begin{split}
{\mathcal P}_l &= {\rm LCM}
\Bigl(1,\frac{\det F}{\det F[1]}, \frac{\det F}{\det F[2]}, 
\ldots, \frac{\det F}{\det F[s]}
\Bigr)\\
&= {\rm LCM}
\left(1,\frac{\det F}{\det F[1]}, \frac{\det F}{\det F[2]-\det F[1]}, 
\ldots, 
\frac{\det F}{\det F[t+1]-\det F[t]}\right) \\
&= {\rm LCM}\left(1, \bigcup_{k=0}^{t}{}^\prime 
\frac{p_{i_{k+1}}p_{i_k}}
{(i_{k+1}-i_k)p_{i_s}}\right),
\end{split}
\end{equation}
where $0 \le t \le s-1$ is the maximum integer 
such that $i_{t+1}>i_t$.

\section{Dynamical period}\label{sec:4}

In this section we shall exclusively consider 
$A^{(1)}_n$ case although the parallel
results are expected for $D^{(1)}_n$.
When $l \rightarrow \infty$, 
one puts $i_k = J_k$ and $t=s-1$ 
in the formula (\ref{eq:lcmsl2}).
Eventually the resulting expression coincides with eq.(4.24) in \cite{YYT},
which gives the period of generic paths in the 
periodic box-ball system containing $m_j$ solitons 
of length $j$.
Here by generic is meant the absence of an `effective 
translational symmetry' \cite{YYT}.
In the present framework it 
corresponds to the time evolution $T_{l = \infty}$ of 
the basic periodic $A^{(1)}_1$ automaton, 
i.e., $\gehn=A^{(1)}_1, \nu^{(a)}_j = L\delta_{j1}$.

To generalize such a connection, 
we invoke the combinatorial version of 
the Bethe ansatz explored in \cite{KKR, KR}.
Given a highest path, namely an element 
$p \in B$ such that 
${\tilde e}_ip=0$ for $1\! \le \! i \!\le \! n$,
one can bijectively attach the data 
$(m^{(a)},r^{(a)})_{a=1}^n$ called 
{\em rigged configuration}.
Here $m^{(a)}=(m^{(a)}_j)$ is a Young diagram 
involving $m^{(a)}_j$ rows of length $j$, and 
$r^{(a)}=(r^{(a)}_j)$ stands for an array of partitions
attached to each `cliff' of $m^{(a)}$.
$\vert m^{(a)} \vert$ is equal to the number of letters 
$a\!+\!1, a\!+\!2, \ldots, n\!+\!1$ contained 
in the corresponding path $p$.
The separate data $(m^{(1)},\ldots, m^{(n)})$ 
and $(r^{(1)}, \ldots, r^{(n)})$ are called 
{\em configuration} and {\em rigging}, respectively.
They obey a special selection rule 
originating in the string hypothesis.
Namely, the $p^{(a)}_j$ defined by (\ref{eq:vacancy}) 
must be nonnegative and the maximum part of 
the partition $r^{(a)}_j$ is not greater than $p^{(a)}_j$
for any $(a,j) \in H$.
It is known that $\det F > 0$ (\cite{KN} Lemma 3.7) for any 
configuration.
Time evolutions of a highest path is not highest in general.
Let $P_h(m) \subseteq B$ be the set of highest paths 
whose configuration is $m=(m^{(1)}, \ldots m^{(n)})$. 

\vspace{0.2cm}\noindent
{\bf Conjecture 1}.
For a highest path $p \in P_h(m)$, 
suppose that 
$T_l^k(p)$ exists for any $k \in \Z_{\ge 1}$.
Then the dynamical 
period of $p$ (minimum positive integer $k$ such that 
$T^k_l(p) = p$) is equal to 
${\mathcal P}_l$ (\ref{lcm})
generically, and its divisor otherwise.

\vspace{0.2cm}
Naturally we expect $\Lambda_l^{\mathcal{P}_l} = 1$, which can 
indeed be verified for $A^{(1)}_1$.
Conjecture 1 implies that the generic period is a function of 
the configuration only and does not depend on the rigging.
We have abruptly combined the Bethe ansatz results in two different regimes.
The first one in section \ref{sec:3} is 
at $q=0$ \cite{KN}, whereas the second one explained here
is relevant to $q=1$ \cite{KKR, KR}.
Conjecture 1
has been confirmed for all the highest paths in 
$B^{\otimes L}_1$ with $L \le 9$ and 
all the $sl_{n \le 4}$ highest paths in 
$B_{l_1}\otimes \cdots \otimes B_{l_L}$ 
with $l_1 + \cdots + l_L \le 7$.
It was observed that non-generic cases are pretty few and 
$T_l^k(p)\,(k\!\ge \!1)$ exists for any highest $p$
in case $B = B^{\otimes L}_1$.

Let us present a few examples of Conjecture 1.
To save the space 
$12\otimes 224$ is written as $12 \cdot 224$ etc.,
and furthermore, $\cdot$ is totally dropped 
for the basic periodic automata.
In each table, the period under 
$T_l$ with maximum $l$ is equal to that under $T_\infty$.
The last table is a preliminary report on the $D^{(1)}_n$ case.

\vspace{0.3cm}\noindent
$A^{(1)}_1$ 
path $= 1 2 1 1 1 2 2 1 2 2 1 1 1 2 2 1 1 2 2$, 
\; configuration $= ((32211))$

\def\arraystretch{1.2}

\begin{tabular}{c|cccc|c}
$l$ & \multicolumn{4}{|c|}{LCM of} & = period \\ \hline
1 & 1,& 19, & 19,& 19 & 19 \\
2 & 1,& 57, & $\frac{171}{22}$, & $\frac{171}{22}$ & 171 \\
3 & 1,& 171,& $\frac{513}{22}$, & $\frac{513}{193}$ & 513 \\
\end{tabular}

\vspace{0.23cm}\noindent
$A^{(1)}_1$ 
path $= 11 \cdot 1112 \cdot 2 \cdot 112 \cdot 122 
\cdot 2 \cdot 2 \cdot 1$, 
\; configuration $= ((43))$

\begin{tabular}{c|ccc|c}
$l$ & \multicolumn{3}{|c|}{LCM of} &  = period \\ \hline
1 & 1,&$ \frac{27}{2}$,& 18 & 54 \\
2 & 1,&$ \frac{27}{4}$,& 9 & 27 \\
3 & 1,&$ \frac{9}{2}$, &6 & 18 \\
4 & 1,& 9,& 3 & 9
\end{tabular}

\vspace{0.23cm}\noindent
$A^{(1)}_2$ 
path $=1 2 1 1 2 1 2 1 3 3 2 2 1 1 1 1 3 3 2 1 1$,
\; configuration = ((43111),(4))

\begin{tabular}{c|ccccc|c}
$l$ & \multicolumn{5}{|c|}{LCM of} &  = period \\ \hline
1 & 1,& 21,& 21,& 21,& 21 & 21 \\

2 & 1, &$\frac{822}{29},$&$\frac{822}{95},$&$\frac{411}{46},$&$\frac{411}{37}$& 822 \\
3 & 1,  &$\frac{959}{22},$& $\frac{959}{176}, $&$\frac{959}{169}, $&$\frac{959}{127}$ & 959\\
4 & 1, & $\frac{2877}{50}, $&$\frac{2877}{400}, $&$\frac{2877}{820}, $&$\frac{2877}{463}$ 
& 2877
\end{tabular}

\vspace{0.23cm}\noindent
$A^{(1)}_3$ path $= 1 \cdot 12 \cdot 3 \cdot 114 \cdot 1 \cdot 2 \cdot 22$, 
\; configuration $= ((3111),(11),(1))$

\nopagebreak

\begin{tabular}{c|ccccc|c}
$l$ & \multicolumn{5}{|c|}{LCM of}  &  = period \\ \hline
1 & 1,&$ \frac{29}{5}$,& 29,&$ \frac{29}{4}$,&$ \frac{29}{4}$ & 29 \\
2 & 1,&  $\frac{58}{7}$,& $\frac{58}{13}$,& $\frac{116}{17}$,& $\frac{116}{17}$ & 116 \\
3 & 1,& $\frac{29}{2}$,& $\frac{29}{12}$,& $\frac{58}{9}$,& $\frac{58}{9}$ & 58 
\end{tabular}

\vspace{0.23cm}\noindent
$D^{(1)}_4$ path $= 1 \cdot 12 \cdot 1 \cdot 223 \cdot 4\bar{2} \cdot 23 \cdot 1$, 
\; configuration $= ((431),(32),(2),(1))$

\begin{tabular}{c|cccccccc|c}
$l$ & \multicolumn{8}{|c|}{LCM of}  &  = period \\ \hline
1 & 1,& $\frac{39}{7}$,&$ \frac{234}{17}$,&$ \frac{234}{17}$,&$ \frac{26}{3}$, &
$\frac{234}{17}$,&$ \frac{117}{11}$, &$\frac{117}{11}$ & 234 \\
2 & 1,& $\frac{39}{5}$, &$\frac{117}{20}$, &$\frac{117}{20}$,&$ \frac{13}{2}$, 
&$\frac{117}{20}$, &$\frac{117}{19}$, &$\frac{117}{19}$ & 117 \\
3 & 1,& 13,&$ \frac{26}{7}$,&$ \frac{26}{7}$, &$\frac{26}{5}$, &$\frac{26}{7}$,&$ \frac{13}{3}$, &$\frac{13}{3}$ & 26 \\
4 &  1,& $\frac{39}{2}$, &$\frac{468}{71}$, &$\frac{468}{227}$, &$\frac{52}{11}$, &$\frac{468}{149}$, &$\frac{117}{31}$, &$\frac{117}{31}$ & 468
\end{tabular}

\vspace{0.15cm}
For instance in the third example, 
configuration $=((43111),(4))$ means that 
$m^{(1)}_1=3, m^{(1)}_3=m^{(1)}_4=1, m^{(2)}_4=1$ and all the 
other $m^{(a)}_j$'s are $0$.

\section{Size of orbit}\label{sec:5}

In this section we only consider the basic periodic $A^{(1)}_n$ automaton, 
i.e., $B=B^{\otimes L}_1$.
In addition to the period under the time evolutions, 
the Bethe ansatz at $q=0$ 
also leads to a formula for the size of certain orbits 
in the periodic automaton.
Recall the quantity 
\begin{equation}\label{eq:R}
\Omega_L(m) = \det F \prod_{(a,j) \in H}
\frac{1}{m^{(a)}_j}
\binom{p^{(a)}_j+ m^{(a)}_j - 1}{m^{(a)}_j - 1} \quad \in \Z
\end{equation}
obtained in \cite{KN} eq.(3.2) 
(denoted by $R(\nu, N)$ therein) as the number of 
off-diagonal solutions to the string centre equation.
Here $\binom{s}{t} = s(s-1)\cdots(s-t+1)/t!$, and 
$L$ and $m=(m^{(a)}_j)$ enter the 
right hand side through (\ref{eq:vacancy}) and (\ref{F}) 
with $\nu^{(a)}_j = L\delta_{a1}\delta_{j1}$.
In this special case the general identity 
known as the combinatorial completeness of the Bethe ansatz at $q=0$ 
(\cite{KN} Corollary 5.6) reads
\begin{equation}\label{eq:comp}
(x_1+ \cdots + x_{n+1})^L 
= \sum_{m}
\Omega_L(m) x_1^{L-q_1}x_2^{q_1-q_2}\cdots x_n^{q_{n-1}-q_n}x_{n+1}^{q_n},
\quad (q_a = \sum_{j\ge 1}jm^{(a)}_j).
\end{equation}
The left hand side is the character of 
$B = B_1^{\otimes L}$. The sum extending over 
all $m^{(a)}_j \in \Z_{\ge 0}$ cancels out except leaving  
the nonzero contributions exactly when 
$L \ge q_1 \ge \cdots \ge q_{n}$.
For example when $n=2$, one has
$\Omega_6(((3),(1)))\!=\!6, \,\Omega_6(((21),(1)))\!=\!36,\,
\Omega_6(((111),(1)))\!=\!18$ summing up to 
$\binom{6}{3,2,1}\!=\!60$ for $(q_1,q_2)\!=\!(3,1)$, whereas 
$\Omega_6(((1),(3)))\!=\!6, \,\Omega_6(((1),(21)))\!=\!-18, \,
\Omega_6(((1),(111)))\!=\!12$ cancelling out for $(q_1,q_2)\!=\!(1,3)$.
In this sense $\Omega_L(m)$ gives a decomposition of the multinomial 
coefficients according to the string pattern $m$.
It is known (\cite{KN} Lemma 3.7) that $\Omega_L(m) \in \Z_{\ge 1}$ 
for any configuration, namely under the condition 
$p^{(a)}_j \ge 0$ for all $(a,j) \in H$.
Moreover it was pointed out in \cite{KOTY} that 
the expression (\ref{eq:R}) for $A^{(1)}_1$ 
simplified by (\ref{Fsl2}) coincides exactly with 
eq.(2.3) in \cite{YYT}, which is the number of
automaton states that contain $m^{(1)}_j$ solitons with length $j$. 
Thus it is natural to ask what is being counted by 
(\ref{eq:R}) for the basic periodic $A^{(1)}_n$ automaton in general.

To deal with this problem we need to consider a more 
general class of time evolutions.
Let $B^{a,j}$ be the crystal of the Kirillov-Reshetikhin module 
$W^{(a)}_j$ \cite{KMN}.
The crystal so far written as $B_l$ is $B^{1,l}$ in this notation.
Given a path $p \in B=B_1^{\otimes L}$,  seek 
an element $v^{a,j} \in B^{a,j}$ such that 
$v^{a,j} \otimes p \simeq p' \otimes v^{a,j}$ for some $p'$ 
under the isomorphism $B^{a,j}\otimes B \simeq B \otimes B^{a,j}$.
If such a $v^{a,j}$ exists and 
$p'$ is unique even when $v^{a,j}$ is not unique, 
we denote the $p' \in B$ by $T^{(a)}_j(p)$.
Otherwise we say that 
$T^{(a)}_j(p)$ does not exist.
We call $p$ {\em evolvable} if $T^{(a)}_j(p)$ exists for all 
members of ${\mathcal T}:= 
\{T^{(a)}_j\mid 1\!\le\! a \!\le\! n,  j \in \Z_{\ge 1}\}$.
For such a $p$ write ${\mathcal T}p = \bigcup_{a,j}T^{(a)}_j(p)$.
We say that $p$ is {\em cyclic} if all the paths 
$p, {\mathcal T}p, {\mathcal T}^2p, \ldots$ are evolvable.
These paths form an orbit
${\rm Orb}(p):= \bigcup_{t\ge 0} {\mathcal T}^t(p)$, 
which is necessarily a finite subset in $B$.
As in the previous section, we 
let $P_h(m) \subseteq B=B_1^{\otimes L}$ denote the set of highest paths 
whose configuration is $m=(m^{(1)}, \ldots m^{(n)})$. 
(Thus $P_h(m)$ is dependent on $L$.)

\vspace{0.2cm}\noindent
{\bf Conjecture 2}.
Given a configuration $m=(m^{(1)}, \ldots, m^{(n)})$, 
one has two alternatives; all the paths 
in $P_h(m)$ are cyclic, or all the paths are not cyclic.
In the former case, the following formula is valid:
\begin{equation}\label{eq:omega}
\Omega_L(m) =\vert \bigcup_{p \in P_h(m)}
{\rm Orb}(p)\vert.
\end{equation}

\vspace{0.2cm}
All the highest paths with length $L \le 5$ are cyclic.
The smallest example of non-cyclic $P_h(m)$ 
emerges at $L=6$, which is $P_h(((22),(2)))$ only.
It consists of the unique highest path 
$p=112233$, which is evolvable but not cyclic.
In fact one has $T^{(2)}_1(p)=213213$ but in the next step 
$[13]\otimes (213213) \simeq (311223)\otimes [13]$ whereas
$[23]\otimes (213213) \simeq (223311)\otimes [23]$. Here 
$[13]\in B^{2,1}$ stands for the column tableau of depth 2, etc.
Thus $p'$ in the above sense is not unique, meaning that 
$213213 \in {\mathcal T}p$ is not evolvable hence $p$ is not cyclic.
For $L=7$, again $P_h(((22),(2)))$ is the unique case
consisting of non-cyclic paths. 
We have checked the conjecture up to $L=8$, where 
there are 5 non-cyclic ones out of 56 possible configurations.
Some examples of Conjecture 2 are presented in the following table.
(We write ${\rm Orb}(m) = \bigcup_{p \in P_h(m)}{\rm Orb}(p)$, 
which also depends on $L$.)

\vspace{0.2cm}
\begin{center}
\begin{tabular}{c|c|c}
$L$ & $m$ & $\Omega_L(m) 
= \vert{\rm Orb}(m)\vert$ \\ \hline
6 & $((3))$ & 6 \\
6 & $((21),(1))$ & 36\\
6 & $((1111),(11),(1))$ & 12 \\
7 & $((31),(1))$ & 56 \\
7 & $((221),(21),(1))$ & 63 \\
7 & $((2111),(21),(1))$ & 133\\
7 & $((2111),(111),(11),(1))$ & 112\\
8 & $((111111),(1111),(11))$ & 4 \\
8 & $((2211),(211),(11),(1))$ & 192\\
8 & $((21111),(211),(11),(1))$ & 304
\end{tabular}
\end{center}

\vspace{0.2cm}
For example in the third case $L=6, m=((1111),(11),(1))$, one has
\begin{equation}
\begin{split}
P_h(m) &= \{121234, 123124, 123412 \},\\
{\rm Orb}(m) &= \{
 121234 ,
 123124 ,
 123412 ,
 124123 ,
 212341 ,
 231241 ,\\
& \qquad \!
 234121 ,
 241231 ,
 312412 ,
 341212 ,
 412123 ,
 412312 
\}.
\end{split}
\end{equation}

\section{Summary}
In this paper we have constructed new periodic 
soliton cellular automata and studied them by a 
novel application of the Bethe ansatz.
Section \ref{sec:2} contains the definition of 
the periodic automata 
associated with any non-exceptional affine Lie algebra $\geh_n$.
Local states range over the crystal $B_l$ of $\geh_n$ and 
the time evolution (\ref{eq:tinf}) is a translation 
in the extended affine Weyl group.
In Section \ref{sec:3}, we have shown that 
Bethe eigenvalues at $q=0$ become 
a $2{\mathcal P}_l$-th root of unity, where 
${\mathcal P}_l$ is explicitly given by
the formula (\ref{lcm}).   
In Section \ref{sec:4},
${\mathcal P}_l$ is conjectured to yield 
the dynamical period of the $A^{(1)}_n$ automata
if $F$ and $F[b,k]$ in (\ref{lcm}) 
are specified by the combinatorial Bethe ansatz \cite{KKR,KR}.  
In Section \ref{sec:5}, the Bethe ansatz character formula 
(\ref{eq:comp}) is found to measure the size of orbits of the automata 
as in Conjecture 2.

\vspace{0.3cm}\noindent
{\bf Acknowledgments} \hspace{0.1cm}
A.K. thanks Mo-Lin Ge, Chengming Bai and 
the organizing committee of DGMTP XXIII for the hospitality 
at Nankai Institute of Mathematics, where a part of this work 
was presented.
He also thanks Masato Okado, Reiho Sakamoto, Taichiro Takagi and 
Yasuhiko Yamada for useful discussion on related topics.
This work is partially supported by Grand-in-Aid for Scientific 
Research JSPS No.15540363.


\begin{thebibliography}{A}

\bibitem [B]{B}
R.~J.~Baxter,
Exactly solved models in statistical mechanics, Academic Press,
London (1982).


\bibitem [Be]{Be}
H.~A.~Bethe,
{Zur Theorie der Metalle, I. Eigenwerte und
Eigenfunktionen der linearen Atomkette},
Z. Physik {\bf 71} (1931) 205--231.

\bibitem[FOY]{FOY}
K.~Fukuda, M.~Okado, Y.~Yamada, 
{Energy functions in box ball systems},
Int.\ J.\ Mod.\ Phys.\ A {\bf 15} (2000) 1379--1392.

\bibitem[HHIKTT]{HHIKTT}
G. Hatayama, K. Hikami, R. Inoue, A. Kuniba, T. Takagi and T. Tokihiro,
{The $A^{(1)}_M$ Automata related to crystals of symmetric tensors},
J.\ Math.\ Phys.\ {\bf 42} (2001) 274-308.

\bibitem[IKO]{IKO}
R. Inoue, A. Kuniba and M. Okado,
{A quantization of box-ball systems},
Rev. Math. Phys. {\bf 16} (2004) 1227--1258.

\bibitem[K]{K}
M.~Kashiwara,
{The crystal base and Littelmann's refined Demazure character formula},
Duke Math.\ J. {\bf 71} (1993)  839--858.

\bibitem[HKOTY]{HKOTY}
G. Hatayama,  A. Kuniba, M. Okado,  T. Takagi and Y. Yamada,
{Scattering rules in soliton cellular automata associated with 
crystal bases}, 
Contemporary Math.\ {\bf 297} (2002) 151--182.

\bibitem[HKT1]{HKT1}
G.~Hatayama, A.~Kuniba, and T.~Takagi,
{Soliton cellular automata associated with crystal bases},
Nucl. Phys. B{\bf 577}[PM] (2000) 619--645.

\bibitem[HKT2]{HKT2}
G.~Hatayama, A.~Kuniba, and T.~Takagi,
{Factorization of combinatorial $R$ matrices and associated
cellular automata},
J. Stat. Phys. {\bf 102} (2001) 843--863.

\bibitem[KNY]{KNY}
K.~Kajiwara, M.~Noumi and Y.~Yamada,
{Discrete dynamical systems with $W(A^{(1)}_{m-1}\times 
A^{(1)}_{n-1})$ symmetry},
Lett. Math. Phys. {\bf 60} (2002) 211--219.

\bibitem[KKM]{KKM}
S-J.~Kang, M.~Kashiwara and K.~C.~Misra,
{Crystal bases of Verma modules for quantum affine Lie algebras},
Compositio Math. {\bf 92} (1994) 299--325.

\bibitem[KMN]{KMN}
S-J.~Kang, M.~Kashiwara, K.~C.~Misra, 
T.~Miwa, T.~Nakashima and A.~Nakayashiki,
{Perfect crystals of quantum affine Lie algebras},
Duke Math.\ J.\ {\bf 68} (1992) 499--607.


\bibitem[KKR]{KKR}
S.V.Kerov, A.N.Kirillov and N.Yu.Reshetikhin,
{Combinatorics, Bethe ansatz, and representations of the 
symmetric group},
Zap. Nauch. Semin. LOMI. \ {\bf 155} (1986) 50--64.

\bibitem[KR]{KR}
A.~N.~Kirillov and N.~Yu.~Reshetikhin,
{The Bethe ansatz and the combinatorics of Young tableaux},
J. Sov. Math. {\bf 41} (1988) 925--955.

\bibitem[KN]{KN} A. Kuniba and T. Nakanishi,
{The Bethe equation at $q=0$, the M\"obius
inversion formula, and weight multiplicities: 
II. The $X_n$ case},
J. Alg. {\bf 251} (2002) 577--618.

\bibitem[KOTY]{KOTY}
A. Kuniba, M. Okado, T. Takagi and Y. Yamada,
{Vertex operators and partition functions 
in the box-ball systems (Japanese)}, 
RIMS K\^oky\^uroku  {\bf 1302} (2003) 91--107.


\bibitem[KOY]{KOY}
A. Kuniba, M. Okado and Y. Yamada,
{Box-ball system with reflecting end}, 
J. Nonlin. Math. Phys.  {\bf 12} (2005) 475--507.


\bibitem[KS]{KS}
A.~Kuniba and J.~Suzuki,
{Analytic Bethe ansatz for fundamental representations of
Yangians}, Commun. Math. Phys. {\bf 173} (1995) 225--264.

\bibitem[MIT]{MIT}
J. Mada, M. Idzumi and T. Tokihiro,
{Path description of conserved quantities of the 
generalized periodic box-ball systems},
J. Math. Phys. {\bf 46} (2005) 022701--19.

\bibitem[R]{R}
N.~Yu.~Reshetikhin,
{The functional equation
method in the theory of exactly soluble quantum systems},
Sov. Phys. JETP {\bf 57} (1983) 691--696.
%
\bibitem[STF]{STF}
E.~K.~ Sklyanin, L.~A.~Takhtajan and L.~D.~Faddeev,
Quantum inverse problem method I. Theor.Math.Phys. {\bf 40} (1980) 688--706.

\bibitem[T]{T} D.~Takahashi,
{On some soliton systems defined by using
boxes and balls}, Proceedings of
the International Symposium on Nonlinear Theory and
Its Applications (NOLTA '93),
(1993) 555--558.


\bibitem[TS]{TS} D. Takahashi and J. Satsuma,
{A soliton cellular automaton},
J. Phys. Soc. Jpn. {\bf 59} (1990) 3514--3519.

\bibitem[TTMS]{TTMS}
T. Tokihiro, D. Takahashi, J. Matsukidaira and J. Satsuma,
{From soliton equations to integrable cellular automata
through a limiting procedure},
Phys. Rev. Lett. {\bf 76},  (1996) 3247--3250.

\bibitem[Y]{Y}
C. N. Yang,
{Some exact results for the many-body problem in one dimension with 
repulsive delta-function interaction},
Phys. Rev. Lett. {\bf 19} (1967) 1312--1314.

\bibitem[YYT]{YYT}
D. Yoshihara, F. Yura and T. Tokihiro,
{Fundamental cycle of a periodic box-ball system},
J. Phys. A: Math. Gen. {\bf 36} (2003) 99--121.

\bibitem[YT]{YT}
F. Yura and T. Tokihiro,
{On a periodic soliton cellular automaton},
J. Phys. A: Math. Gen. {\bf 35} (2002) 3787--3801.
\end{thebibliography}
\end{document}